    \documentclass[twocolumn]{revtex4-2} 
    \usepackage{graphicx}
    \usepackage{subfig}
    \newcommand{\replaced}[2]{#2}
    \newcommand{\deleted}[1]{}
    \newcommand{\added}[1]{#1}

\usepackage{mathtools}
\usepackage{amsmath,amssymb}
\usepackage{siunitx}
\usepackage{upgreek}
\usepackage{url}

\usepackage{tabularx,array,booktabs}
\usepackage{soul}
\usepackage{varioref}
\usepackage{comment}
\usepackage{enumerate}
\usepackage{nicefrac}

\newcommand{\R}{\mathbb R}

\newcommand{\Comsol}{COMSOL Multiphysics\textsuperscript{\textregistered}}

\DeclareMathOperator{\atanh}{atanh}

\begin{document}


\title{Non-dispersive graded impedance acoustic lenses}


\author{Sebastiano Cominelli}
\email[]{cominelli@lma.cnrs-mrs.fr}
\affiliation{Aix-Marseille Université, Laboratoire de Méchanique et d'Acoustique, Marseille, France}


\date{\today}

\begin{abstract}
    Acoustic lenses are typically based on refractive index profiles derived from the geometric approximation of high-frequency waves, yet the critical issue of impedance mismatch is often neglected. Mismatched devices suffer from unwanted reflections and dispersion, which can significantly degrade performance in practical applications.
    In this work, we propose impedance profiles for lenses to achieve efficient wave transmission while maintaining the desired refractive index and minimizing dispersion effects.
    A family of impedance profiles is derived from the acoustic wave equation such that the phase velocity is preserved. First, the 1D setting is considered to explain how dispersion occurs inside a lens and at its interfaces.
    Then, the method is applied to 2D axisymmetric configurations where the impedance mismatch is radially redistributed. These profiles are demonstrated in the acoustic setting of a L\"uneburg lens, but can be easily extended to more general scenarios such as imaging or cloaking in air and water, where matching the impedance of the background poses significant challenges. 
\end{abstract}


\maketitle


\section{Introduction}

Historically, lenses are developed using geometric optics, where wave propagation is approximated by ray trajectories that obey Snell's law. The simplest lenses consist of a homogeneous piece of dielectric whose geometry bends the rays at the interfaces with the background medium. More advanced lenses---such as the well-known Maxwell fisheye \cite{whewell1854cambridge} and the L\"uneburg lens \cite{luneburg1966mathematical}---rely on a spatially varying refractive index to manipulate a wave field. Following these pioneering works, novel index profiles for imaging  have been continuously proposed \cite{minano2006perfect,lin2009gradient,narimanov2009optical,su2017broadband}.
This concept has recently been extended by the so-called transformation of coordinates, leading to more powerful devices such as concentrators, that focus waves within a target region \cite{rahm2008design}, perfect lenses, enabled by negative index profiles \cite{pendry2000negative,ramakrishna2004spherical}, and cloaking, which is emerging as one of the most striking applications \cite{Pendry2006,Leonhardt2006,cummer2007one,Norris2008}.
\\
The realization of such devices typically involves approximating the desired index profile with discrete material layers, called graded index (GRIN) materials.
The advent of architected crystals has revolutionized this approach and extended graded index concepts to other fields, including acoustics. In this field, phononic crystals play a key role in providing the building blocks for acoustic devices \cite{laudebook}.

Achieving a specific index profile does not require unique material properties, as it depends only on the wave speed, which is a function of, for example, dielectric permittivity and magnetic permeability in electromagnetism, or mass density and bulk modulus in acoustics. However, optimal performance is achieved when energy is efficiently transferred through the device, which occurs when the impedance of the device is matched to the surrounding medium.
\added{
Over the past two decades, numerous acoustic devices have been developed by adapting electromagnetism concepts to acoustics \cite{cummer2007one,Norris2008}. A key advantage is that the speed of sound is not fundamentally limited, unlike the speed of light. Moreover, while modifying magnetic permeability is difficult, acoustic properties such as density and bulk modulus are more easily tailored \cite{laudebook,su2017broadband}--particularly with the advent of advanced additive manufacturing technologies \cite{kadic2014pentamode}.
This is particularly evident in underwater acoustics, where significant progress has been made with two-dimensional systems \cite{su2017broadband,chen2017broadband,quadrelli2021experimental}. However, extending these approaches to three-dimensional devices remains a substantial challenge. The difficulty of achieving impedance matching with the surrounding medium \cite{li2019three,brambilla2024high} has limited the number of experimental realizations of 3D devices \cite{bi2018experimental,allam20203d}, which are nonetheless impedance-mismatched.
}
The challenge of impedance matching has been studied in several areas, such as electromagnetic transmission lines \cite{ramo1994fields}, acoustic horns \cite{webster1919acoustical,klopfenstein1956transmission,benade1974plane}, \replaced{and mechanical interfaces}{
medical ultrasound focusing \cite{pedersen1982impedance,li2017broadband},
impact shock mitigation \cite{hui2011new}, sound transmission \cite{fleury2014metamaterial}, and the recent usage of metamaterials have increased the possibilities to obtain perfect matching devices \cite{luo2016ultratransparent,li2017broadband,hui2011new}.
}
A key goal in these studies is to achieve near-perfect transmission over specific frequency intervals. In the one-dimensional (1D) setting, two media of different impedance are connected by a slab whose properties must be designed.
Two well-established configurations allow efficient wave transmission: the quarter-wavelength constant impedance matching plate and the exponentially graded impedance \cite{ramo1994fields,pedersen1982impedance}. The latter represents the continuous limit of the former as the number of discrete matching layers increases \cite{kossoff1966effects}.
In these scenarios, low-frequency waves are targeted because the signal wavelength is typically longer than the slab thickness, and only the transmission characteristics from one side of the slab to the other are relevant.
\replaced{
However, these studies are limited to 1D devices and do not take into account the phase distortion of the signal due to the device itself.
\\
The GRIN devices recently studied in the literature often interact with a signal for a few wavelengths \cite{rahm2008design,cai2007optical,Popa2011,quadrelli2021experimental,bi2018experimental,li2019three,cominelli2022design,cominelli2024optimal} because the wavelength is comparable to the device size. Working in such a mid-frequency range is necessary for several reasons, for example in order to exploit the low-frequency effective properties of a grating, or to allow the wave to interact with sensors that measure a spatial average of the wave field.
In these scenarios, how the wave propagates within the device itself becomes important, and a constant impedance mismatch has been used in several contexts to preserve it. For example, this strategy has been used to obtain a solid device that cloaks pressure waves in air \cite{Popa2011,zigoneanu2014three}, and to design a cloak for water waves \cite{dupont2016cloaking} to overcome the impossibility of changing the acceleration of gravity.}
{
This is in contrast with traditional GRIN lenses, which are usually designed under high-frequency (ray) approximations, where smooth impedance variations can be neglected.

The devices recently studied in the literature often operate in an intermediate regime, where the device size spans a few wavelengths \cite{rahm2008design,cai2007optical,Popa2011,quadrelli2021experimental,bi2018experimental,li2019three,cominelli2022design,cominelli2024optimal}. This mid-frequency range is necessary, for example, to exploit the effective low-frequency behavior of periodic gratings.
In this regime, the working frequencies are too low to neglect impedance variations, yet too high for the standard low-frequency transmission problem. 
As a result, efforts have been made to create materials with extreme physical properties to reduce the mismatch \cite{brambilla2024high,kadic2014pentamode}; however, a constant impedance mismatch is often introduced as part of the design strategy.
For example, this approach has been employed to create solid structures that cloak pressure waves in air \cite{Popa2011,zigoneanu2014three}, and to design a cloak for water waves where altering the gravitational acceleration is not an option \cite{dupont2016cloaking}.
}
\\
A constantly mismatched device directs a wave field through it as desired, but at the cost of reflections at the interface with the background. Then, in all of these settings, the impedance is as close as possible to the surrounding medium.
\replaced{Graded impedance}{In this direction, graded impedance could offer} advantages. For example, a smooth transition with the surrounding medium \replaced{can}{could} be achieved, reducing unwanted reflections, and the device properties \replaced{can}{could} be modified more in the areas where the realization is more critical. 
The drawback is that an impedance profile may introduce dispersion, which affects wave transmission in the form of phase distortion.
\added{ The ideal scenario is to modify a device’s impedance without compromising its ability to transmit the signal effectively. A key requirement is to preserve the signal integrity by ensuring that all frequency components propagate at the same speed as they would in an ideally matched medium, thereby maintaining the correct arrival time. Devices that achieve this are referred to as non-dispersive. This is of major importance in two or three dimensions where even small phase changes can cause destructive interference between different propagation paths, resulting in a significant intensity loss.
}
\\
In this letter, we propose a method for changing the impedance of a device while preserving its index profile and its non-dispersive properties.
\added{An impedance profile is derived which is non-dispersive within the lens, but introduces dispersion at the lens boundaries, especially for low-frequency waves.}
The method is first introduced in the 1D scenario to get a close insight. We compute the reflection and transmission coefficients when the impedance has a non-continuous derivative and \replaced{derive}{obtain} a formula that is more general than the canonical jump impedance transmission.
\deleted{An impedance profile without dispersion within the lens is derived.}
We extend the method to 2D axisymmetric devices where the mismatch is redistributed along the radius. The case of the L\"uneburg lens is considered as an example.
This approach facilitates practical implementation and reduces the drawbacks of a mismatched device for mid and high frequencies.
\\
The paper focuses on the acoustic framework because it remains a challenge to mimic the impedance of typical acoustic media. For example, creating materials as light as air \cite{Popa2011,zigoneanu2014three,kan2015broadband} or with the high bulk modulus of water \cite{brambilla2024high,li2019three,bi2018experimental} is not trivial.
Also, mass density and bulk modulus are usually adjusted independently when designing phononic crystals \cite{kadic2014pentamode,li2019three,cominelli2022design,cominelli2024optimal,brambilla2024high}.
However, the theory is easily extended to other physical systems governed by similar mathematical structures, such as polarized electromagnetic waves or anti-plane elasticity, provided that the stiffness- and density-related terms can be tuned independently.


\section{Improved transmission through graded mismatch}

\begin{figure}
    \centering
    \includegraphics[width=0.75\linewidth]{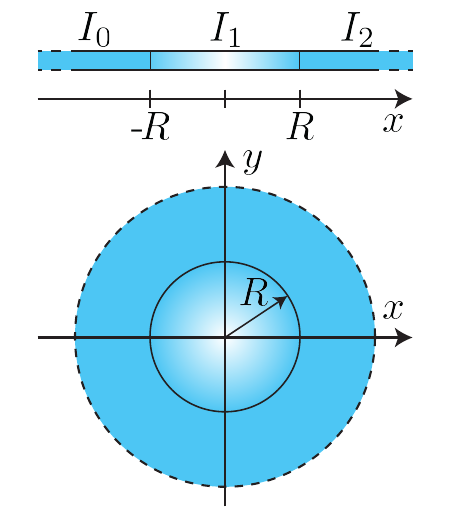}
    \caption{Schematic of a 1D and a 2D lens with radius $R$.}
    \label{fig:1D e 2D lenses}
\end{figure}
Consider the one-dimensional (1D) lens schematized in Figure~\ref{fig:1D e 2D lenses}, where the space is divided into three intervals $I_0 = (-\infty, -R)$, $I_1 = [-R, R]$, and $I_2=(R,+\infty)$ along the $x$-axis, $R\in\R^+$. We assume constant properties in $y$ and $z$ directions and consider waves propagating only along $x$. The same homogeneous medium occupies $I_0$ and $I_2$. Its mass density and bulk modulus are $\rho_0$ and $K_0$, respectively.
Conversely, the properties $\rho_1$ and $K_1$ of medium 1, the lens, are graded. The acoustic properties are typically also expressed in terms of acoustic impedance $Z_i\coloneqq\sqrt{\rho_iK_i}$ and phase velocity $c_i\coloneqq\sqrt{K_i/\rho_i}$.
In this arrangement, a signal passing through the lens experiences amplitude and phase distortion according to the lens properties.
Assuming a lens matched with the environment (i.e.\ $Z_1=Z_0$), only a phase shift occurs due to the velocity profile defined as $n_1(x)\coloneqq c_0/c_1(x)$, that is usually designed to achieve imaging---typically in 2D or 3D lenses.
We look for a clever way to grade the impedance of the lens such that the phase shift is preserved as much as possible.
For the sake of simplicity, we choose the most simple lens $n_1(x)\equiv 1$, but the method applies to the general case as discussed below for a 2D lens.
Using the properties of medium 0 as reference, we write
\begin{gather}
    \frac{K_0}{\rho_0} = \frac{K_1}{\rho_1},
    \qquad
    \frac{\rho_1}{\rho_0} = \frac{K_1}{K_0} =
    \frac{Z_1}{Z_0} = \alpha(x),
\end{gather}
where $\alpha(x)$ is a bounded arbitrary function $\alpha\colon I_1\to\R^+$ that grades the impedance in the medium 1.

The wave equation inside $I_1$ reads as
\begin{equation} \label{eq:wave 1D}
    \Big(\frac{1}{\alpha\rho_0} p_x\Big)_x = \frac{1}{\alpha K_0} p_{tt},
\end{equation}
where the subscripts $x$ and $t$ indicates the partial derivatives in space and time, respectively.
Given a graded impedance $\alpha$, waves propagate differently through $I_1$ , which can result in dispersion.
To gain insight into the problem, we examine the well-known scenario of the exponentially graded impedance $\alpha_{\text e} = a_{0,\text e}e^{2a_{1,\text e} x}$ \cite{pedersen1982impedance}. In this case, \eqref{eq:wave 1D} admits solutions in the form
\begin{equation}\label{eq:p_alpha exp}
    p_\alpha = e^{a_{1,\text e} x}\,\Re\{e^{j\omega t} (a_{2,\text e} e^{j\kappa(\omega)x} + a_{3,\text e} e^{-j\kappa(\omega)x} )\},
\end{equation}
where the wavenumber $\kappa(\omega)$ satisfies the dispersion equation $\kappa^2=\omega^2/c_0^2 - a_{1,\text e}^2$, and $a_{0,\text e},\,a_{1,\text e},\,a_{2,\text e},\,a_{3,\text e}\in\R$ are arbitrary constants. $j$ is the imaginary unit and $\Re$ the real-part operator.
The medium is clearly dispersive since the phase velocity is frequency dependent:
\begin{equation}\label{eq:exp disp}
    c(\omega) \coloneqq \frac{\omega}{\kappa(\omega)} = c_0 \frac{\omega}{\sqrt{\omega^2 - a_{1,\text e}^2c_0^2}}.
\end{equation}
In addition, for $\omega < \omega_{\text e} \coloneqq |a_{1,\text e}| c_0$ the wavenumber becomes purely imaginary, resulting in evanescent waves. Thus, when the frequency is below $f_\text{e}=\frac{\omega_\text{e}}{2\pi}$, the graded medium does not efficiently transmit energy. This behavior can be qualitatively understood by considering that the wavelength at low-frequency is large compared to the spatial variation of the medium properties, and a continuous reflection occurs. Conversely, for high frequency ($\omega\gg\omega_{\text{e}}$) the reference sound speed $c_0$ is recovered.

\added{
Using the Euler's equation $\rho\dot{u}=-p_x$, the particle velocity $u$ is:}
\begin{multline}\label{eq:u_alpha exp}
    u = \frac{a_{3,\text e}}{a_{0,\text e}\rho_0} e^{-a_{1,\text e}x}
    \,\Re\Big\{\frac{e^{j\omega t}}{\omega}\big[
    \big(ja_{1,\text e}-k(\omega)\big) e^{jk(\omega)x}
    \\+
    \big(ja_{1,\text e}+k(\omega)\big) e^{-jk(\omega)x}
    \big]\Big\}.
\end{multline}\added{
The role of $ja_{1,\text e}$ is clear when, e.g., the right-propagating wave of \eqref{eq:u_alpha exp} is written as}
\begin{equation}
    \frac{a_{3,\text e}}{a_{0,\text e}\rho_0} e^{-a_{1,\text e}x}
    \,\Re\Big\{
    \frac{ja_{1,\text e}}{\omega}e^{j(\omega t - k(\omega)x)}
    +\frac{k(\omega)}{\omega} e^{j(\omega t -k(\omega)x)} 
    \Big\}.
\end{equation}\added{
which shows that, differently from a homogeneous medium, the impedance profile induces two waves that travels at the same speed $c(\omega)$, but with different amplitude. In particular, the amplitudes $ja_{1,\text e}/\omega\to0$ and  $k(\omega)/\omega\to1/c_0$ as frequency increases.
}\\
When $\omega>\omega_{\text{e}}$, the exponential amplification $e^{\pm a_1 x}$ in \eqref{eq:p_alpha exp} \added{and \eqref{eq:u_alpha exp}} is the effect of energy conservation of the traveling waves. Indeed, the acoustic potential \added{and kinetic} energy \replaced{density is}{densities are} defined by $w_\text{P} = p^2 / (2K)$ \added{and $w_\text{K} = \rho u^2/2$, respectively}, and for the right propagating wave of \eqref{eq:p_alpha exp} \replaced{it is}{and \eqref{eq:u_alpha exp} they are}
\begin{equation}
\begin{aligned}
    w_\text{P}(x,t) &= 
           \frac{a_{3,\text e}^2}{a_{0,\text e}}  \frac{1}{2K_0}   \Big(\Re\{e^{j(\omega t- \kappa(\omega) x)}\}\Big)^2,
           \\
    w_\text{K}(x,t) &= 
        \frac{a^2_{3,\text e}}{a_{0,\text e}\omega^2} \frac{1}{2\rho_0}   \Big(\Re\big\{  \big(ja_{1,\text e}+\kappa(\omega)\big) \;e^{j(\omega t- \kappa(\omega) x)}\big\}\Big)^2.
\end{aligned}
\end{equation}
Thus, \replaced{$w$}{$w_\text{P}$ and $w_\text{K}$} propagate with the same speed of the wave, \added{and no exponential amplification due to $e^{\pm a_1 x}$ occurs.
}

\subsection{Transmission without dispersion}
If $\alpha$ is constant, Eq.~\eqref{eq:wave 1D} has the well-known solutions $p = p_1(x \pm c_0t)$, representing backward and forward traveling waves with an arbitrary shape that is preserved during the propagation.
Conversely, if the properties vary in space the shape of the wave changes due to \added{dispersion and} energy conservation, as in the example shown above.
In the general case of graded impedance $\alpha$, we wonder whether there exists a property distribution that \added{avoids dispersion and} preserves the shape of any transmitted wave, even though the amplitude is gradually scaled as the wave propagates.
To compute such an impedance profile, we look for a traveling wave in the form
\begin{equation}
    p_\alpha = G\big(x, p_1(x-c_0t)\big),
\end{equation}
that is a wave propagating with speed $c_0$ and whose amplitude changes in space according to the function $G$. By substitution into Eq.~\eqref{eq:wave 1D} it yields:
\begin{multline}\label{eq:G wave 1D}
    \Big[\Big(\frac1\alpha\Big)' G_x + \frac1\alpha G_{xx}\Big]
    + \Big[\frac2\alpha G_{xp} + \Big(\frac1\alpha\Big)' G_p\Big]  p_{1\,x}
    \\
    + \frac{G_{pp}}{\alpha}\Big[( p_{1\,x})^2 - \frac{\rho_0}{K_0} ( p_{1\,t})^2\Big]
    = 0,
\end{multline}
where the identity~\eqref{eq:wave 1D} holding for $p_1$ has been used to simplify the equation. The subscripts $x$ and $p$ applied to $G$ represent the partial derivatives with respect to the two explicit arguments of $G$, respectively, while the prime apex is the total derivative with respect to $x$.
Eq.~\eqref{eq:G wave 1D} is satisfied for all $p_1$ if its three terms cancel out independently:
\begin{subequations}
\begin{align}\label{eq:diff 1D 1}
    \Big(\frac1\alpha\Big)' G_x + \frac{1}{\alpha}G_{xx} = 0,
    \\\label{eq:diff 1D 2}
    \frac2\alpha G_{xp}+\Big(\frac1\alpha\Big)'G_p = 0,
    \\ \label{eq:diff 1D 3}
    \frac{G_{pp}}{\alpha}  = 0.
\end{align}
\end{subequations}
Eq.~\eqref{eq:diff 1D 3} implies $G_{pp} = 0$, then
\begin{align}
    G(x,p)&=g(x)\, p + h(x)
\end{align}
for some functions $g$ and $h$.
This result is expected since it is the only possibility for $G$ to preserve the linearity of the wave equation with respect to the pressure field.
Substituting into~\eqref{eq:diff 1D 1} and \eqref{eq:diff 1D 2} we obtain, respectively:
\begin{subequations}
\begin{gather}
    \Big[\Big(\frac1\alpha\Big)'h'+\frac{h''}{\alpha}\Big]
    + \Big[\Big(\frac1\alpha\Big)' g' + \frac{g''}{\alpha}\Big]\tilde p_1 = 0
    \\
    2\frac{g'}{\alpha} + \Big(\frac1\alpha\Big)' g = 0
\end{gather}
\end{subequations}
that correspond to the three equations
\begin{subequations} \label{eq:diff 1D 456}
\begin{align}\label{eq:diff 1D 4}
    \Big(\frac1\alpha\Big)' h'+\frac{h''}{\alpha} = 0,
    \\ \label{eq:diff 1D 5}
    \Big(\frac1\alpha\Big)' g' + \frac{g''}{\alpha} = 0,
    \\ \label{eq:diff 1D 6}
    2\frac{g'}{\alpha} + \Big(\frac1\alpha\Big)' g = 0.
\end{align}
\end{subequations}
This system of differential equations is solved to find $\alpha$, $g$, and $h$ up to integration constants.
From~\eqref{eq:diff 1D 4} and \eqref{eq:diff 1D 5}:
\begin{align}
    \Big(\frac{h'}{\alpha}\Big)' = 0,
    &&
    \Big(\frac{g'}{\alpha}\Big)' = 0.
\end{align}
Assuming $g\neq0$, Eq.~\eqref{eq:diff 1D 6} is equivalent to
\begin{equation}\label{eq:diff 1D 7}
        \Big(\frac{g^2}{\alpha}\Big)' = 0.
\end{equation}
Finally, the system is solved by
\begin{subequations}\label{eq:def fgh 1D }
\begin{align}
    \alpha_{\text{nd}}(x) &
    = a_{0,\text{nd}}(x+a_{1,\text{nd}})^{-2}, \label{eq:def f 1D}
    \\
    g(x) &
    = a_{2,\text{nd}}(x+a_{1,\text{nd}})^{-1},  \label{eq:def g 1D}
    \\
    h(x) &
    = a_{3,\text{nd}}(x+a_{1,\text{nd}})^{-1}+a_{4,\text{nd}},
\end{align}
\end{subequations}
for some constants $a_{i,\text{nd}}$, $i\in\{0,1,2,3,4\}$ that are determined by setting the problem and imposing the boundary conditions.
Hence, a medium whose properties are graded as the non-dispersive profile \eqref{eq:def f 1D} sustains traveling waves of the type $p_\alpha = g(x)\,p_1(x\pm c_0t) + h(x)$, for any function $p_1$.
Eq.~\eqref{eq:wave 1D} is invariant with respect to a pressure shift of $h$, just as is the homogeneous case for any first-order polynomial.
Note that the homogeneous case is recovered when both $\alpha$ and $g$ are constant (i.e.\ $a_{0,\text{nd}}\propto a_{1,\text{nd}}^2\propto a_{2,\text{nd}}^2\to\infty$).

\added{The particle velocity can now be computed in time domain:}
\begin{align}
    u 
    &= \frac{\pm1}{c_0\rho_0\alpha(x)}\Big(g(x)\,p_1(x\pm c_0t) + g'(x)P_1(x\pm c_0t)\Big)  \\
    & = \frac{\pm a_{2,\text{nd}}}{c_0\rho_0a_{0,\text{nd}}}\Big((x+a_{1,\text{nd}})\,p_1(x\pm c_0t)-P_1(x\pm c_0t)\Big),
\end{align}\added{
where $P_1\coloneqq\int_0^{c_0t}p_1(\xi)\text d\xi$.
Thus, the non-dispersive profile induces two waves traveling at the same speed $c_0$. There are two differences with respect to the exponential profile: (i) the velocity of both waves does not depend on frequency and (ii) only one wave ($p_1$) is modulated in space by $(x+a_{1,\text{nd}})$.}

\replaced{A major result is that energy is transmitted efficiently for any frequency. Indeed, the acoustic potential energy density is}{Let us define how energy is transmitted; the acoustic energy densities are}
\begin{equation}\label{eq:energy conserv}
\begin{aligned}
    w_\text{P}(x,t) &=  \frac{g(x)^2}{\alpha_{\text{nd}}(x)}  \frac{1}{2K_0} p_1(x\pm c_0t)^2 
    \\
           &= \frac{a_{2,\text{nd}}^2}{a_{0,\text{nd}}}\, w_{\text P,0}(x\pm c_0t),
   \\
    w_\text{K}(x,t)
    &=  \frac{1}{2K_0\,\alpha(x)} \Big(g(x)\,p_1(x\pm c_0t) + g'(x)P_1(x\pm c_0t)\Big)^2
    \\
    & = \frac{a_{2,\text{nd}}^2}{2K_0a_{0,\text{nd}}}\Big(p_1(x\pm c_0t) - (x+a_{1,\text{nd}})^{-1} P_1(x\pm c_0t)\Big)^2,
\end{aligned}
\end{equation}
where $w_{\text P,0}$ is the potential energy of the field $p_1$ propagating in a homogeneous medium; $w_\text{P}$ propagates with the same speed.
\deleted{of the pressure wave, meaning that no dispersion occurs. Note that this property is a direct consequence of \eqref{eq:diff 1D 7}.}
\added{
Conversely, the kinetic energy consists of a constant term $p_1$ and a modulated term proportional to $P_1$. The former corresponds to energy flow in a homogeneous medium (as for $w_\text{P}$), while the latter represents a portion of energy that is not efficiently transported, as it explicitly depends on $x$. Since $P_1$ is the integral of $p_1$, when $p_1$ forms a wave packet with zero mean, $P_1$ vanishes once the packet has passed, indicating that no energy is left behind.
}

For the 1D lens, it is desired that symmetry with respect to $x=0$ is preserved by the impedance grading. This is possible only if $a_{1,\text{nd}}=0$ in \eqref{eq:def f 1D}, i.e.\ $\alpha=a_{0,\text{nd}}x^{-2}$, leading to $\alpha(x\to0)\to\infty$. To avoid this unfeasible choice, we define the function
\begin{equation}\label{eq:alpha abs}
    \tilde\alpha_{\text{nd}}(x)\coloneqq
    \begin{dcases}
        \alpha_-(x) = a_{0,\text{nd}}(x-a_{1,\text{nd}})^{-2} & x<0,
        \\
        \alpha_+(x) = a_{0,\text{nd}}(x+a_{1,\text{nd}})^{-2} & x\ge0.
    \end{dcases}
\end{equation}
So that the lens impedance is graded on both sides by a function of the form of \eqref{eq:def f 1D}, at the cost of further reflections at $x=0$ due to a slope discontinuity.
A similar stratagem is required below to use the exponential profile.

\subsection{Interface between two graded impedance media}
\replaced{Let us now consider what happens}{In the following, we study the case of normal incidence of a wave} at the interface $x=0$ inside medium 1, where the properties are graded by $\alpha_-(x)=\rho_-/\rho_0=K_-/K_0$ for $x<0$, and $\alpha_+(x)=\rho_+/\rho_0 =K_+/K_0$ for $x\ge0$, respectively. If the profile given by \eqref{eq:alpha abs} is adopted\replaced{,}{ we have that} $\alpha_-(-x)=\alpha_+(x)$, but \added{in the following} they are kept independent \added{for} the sake of generality.
Given the functions $\alpha_-$ and $\alpha_+$ in the form of \eqref{eq:def f 1D}, the modulations $g_-$ and $g_+$ are defined through \eqref{eq:def g 1D} up to arbitrary constants that we choose such that $g_-|_0=g_+|_0=1$---to uniquely define the reflected and transmitted amplitude.
Using the time harmonic expansion, the pressure field at $x<0$ is considered as the superposition of an incident wave $p_I=\Re\{g_-(x)\,e^{j(\omega t - \kappa x)}\}$ with unitary amplitude at the interface, and a reflected wave $p_R = \Re\{R\,g_-(x)\, e^{j(\omega t +\kappa x)}\}$, with amplitude $R$.
At $x\ge0$, a right propagating wave is transmitted $p_T = \Re\{T\,g_+(x)\, e^{j(\omega t - \kappa x})\}$. $\omega$ is the angular frequency and $\kappa=\omega/c_0$ the wave number, that is the same on both sides since the sound speed is $c_-=c_+=c_0$ everywhere and for every frequency.
The continuity conditions read as
\begin{equation}
\begin{dcases}
    (p_I + p_R)\big|_{0^-} = p_T\big|_{0^+}
    \\
    \frac{1}{\rho_-}(p_I + p_R)_x\big|_{0^-} = \frac{1}{\rho_+}(p_T)_x\big|_{0^+}
\end{dcases}.
\end{equation}
By substitution, it yields
\begin{equation}\label{eq:generalized TR}
\begin{dcases}
    T = \frac{2 \kappa Z_+}  {\kappa(Z_+ + Z_-) - j (Z_+g'_-|_0 - Z_-g'_+|_0)}
    \\
    R = \frac{\kappa(Z_+-Z_-) + j(Z_+g'_-|_0 - Z_-g'_+|_0)}{\kappa(Z_+ +Z_-) - j (Z_+g'_-|_0 - Z_-g'_+|_0)}
\end{dcases},
\end{equation}
where $Z_\pm = \alpha_\pm|_0 Z_0$. The classical formulae for $T$ and $R$ are recovered when the media are homogeneous, i.e.\ $g'_-=g'_+=0$.
In the general scenario, reflections occur for two reasons: (i) when there is an impedance mismatch ($Z_-\neq Z_+$), or (ii) when the wave changes shape ($g'_-\neq g'_+$). The latter phenomenon is frequency dependent, with reflections being stronger at lower frequencies. In the limit case, the zero frequency component experiences total reflection, with $T\to0$ and $R\to-1$. Conversely, at high frequencies ($\big|\kappa/g'_-|_0\big|,\,\big|\kappa/g'_+|_0\big|\gg1$), only the impedance mismatch at the interface dominates the transmission, giving $T\approx2Z_+/(Z_++Z_-)$ and $R\approx(Z_+-Z_-)/(Z_++Z_-)$, and total transmission occurs when $Z_-=Z_+$. 
\\
Nearby the interface, any impedance profile can be approximated in the form of \eqref{eq:def f 1D}, so the formulae~\eqref{eq:generalized TR} extend to impedance profiles with a jump discontinuity and/or non-continuous derivative, when the refraction index is constant.
In this case, we recall \eqref{eq:diff 1D 7} about the interface and replace $g'_\pm|_0$ by $\nicefrac{1}{2}\, Z'_\pm/Z_\pm$ , where $Z'_\pm \coloneqq \alpha'_\pm|_0\, Z_0$:
\begin{equation}\label{eq:generalized TR 2}
\begin{dcases}
    T = \frac{2\kappa Z_+}  {\kappa(Z_+ + Z_-) - \nicefrac{j}{2} (Z_+ Z'_-/Z_- - Z_- Z'_+/Z_+)}
    \\
    R = \frac{\kappa(Z_+-Z_-) + \nicefrac{j}{2}(Z_+ Z'_-/Z_- - Z_- Z'_+/Z_+)}{\kappa(Z_+ + Z_-) - \nicefrac{j}{2} (Z_+ Z'_-/Z_- - Z_- Z'_+/Z_+)}
\end{dcases}.
\end{equation}


\begin{table}[]
    \centering
    $\begin{array}{c|ccc|cc}
    \alpha_{\text{c}} \,[-] & a_{0,\text e} \,[-] & a_{1,\text e}R \,[-] & f_{\text{e}}  \,[Hz] & a_{0,\text{nd}}R^2 \,[-]& a_{1,\text{nd}}R\,[-]
    \\    \midrule
    0.1 & 0.331 & 0.326 & 0.79      & 0.321 & 0.318
    \\
    0.2 & 2.188 & 2.138 & 0.39       & 2.094 & 2.054
    \end{array}$
    \caption{Coefficients of the impedance profiles.}
    \label{tab:coeff}
\end{table}

\replaced{To have a transmission that does not cause dispersion, one need to grade the impedance using a function in the form of \eqref{eq:def f 1D} and such that $Z_-'/Z_-^2 = Z'_+/Z_+^2$ on any interface. It is easy to show that, only a constant impedance meets both conditions.
Therefore, any non-constant profile adopted to rescale the properties of the 1D lens introduces a frequency-dependent transmission either at the lens-background interfaces or within the lens itself. On top of that, multiple reflections inside the lens deteriorate the performance.}{
Thus, the impedance of a dispersion-free device must vary according to a function of the form \eqref{eq:def f 1D}, and the condition $Z_-' / Z_-^2 = Z_+' / Z_+^2$ must be satisfied at every interface.
Since the surrounding medium typically has constant impedance, we have $Z_-' = 0$. However, by enforcing $Z_+' = 0$ at the surface of the lens it is easy to prove that only a constant impedance profile fits the form \eqref{eq:def f 1D}.
Indeed, a constant impedance profile eliminates dispersion, but it leads to strong reflections over most of the frequency spectrum--except for narrow transmission bands, as shown later. In contrast, non-constant impedance profiles can enhance transmission by reducing mismatch at the interface, but as we have shown, they may introduce dispersion both at the boundaries and within the lens bulk.
An exception is provided by profiles of the form \eqref{eq:def f 1D}, which produce dispersion only at the interfaces while preserving a dispersion-free interior. This highlights that a trade-off between minimizing dispersion and maximizing transmission must be considered in lens design.
In addition, the impact of the imperfect energy transport described earlier remains unclear. In the following, we numerically evaluate which impedance profiles best balance these competing objectives.
}

\subsection{Numerical examples}

\begin{figure*}
    \centering
    \subfloat[]{\includegraphics[scale=0.45]{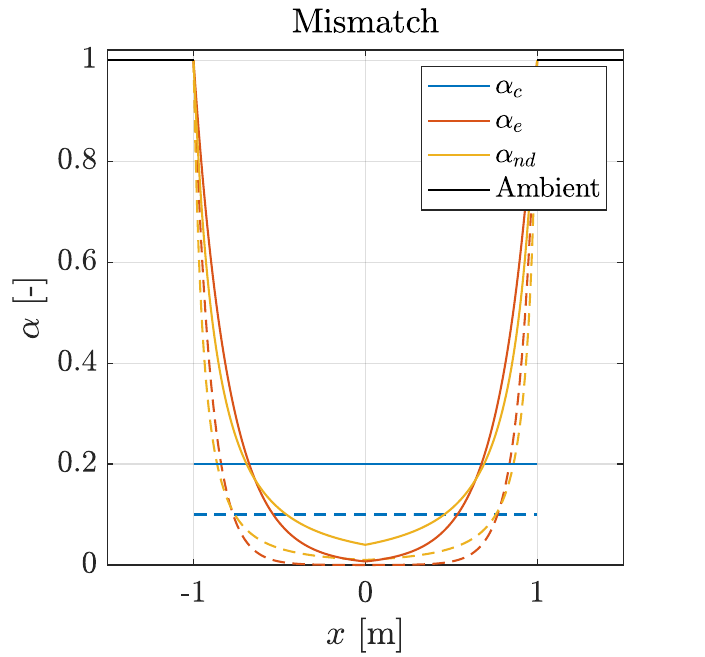}\label{fig:alpha 1D}}
    \subfloat[]{\includegraphics[scale=0.46]{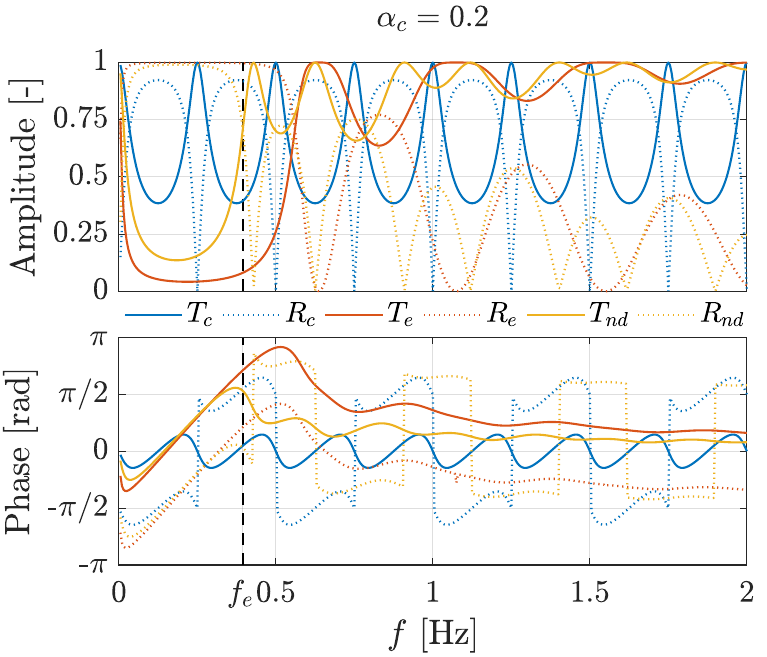}\label{fig:T and R slab-b}}
    \subfloat[]{\includegraphics[trim=30 0 0 0,clip,scale=0.46]{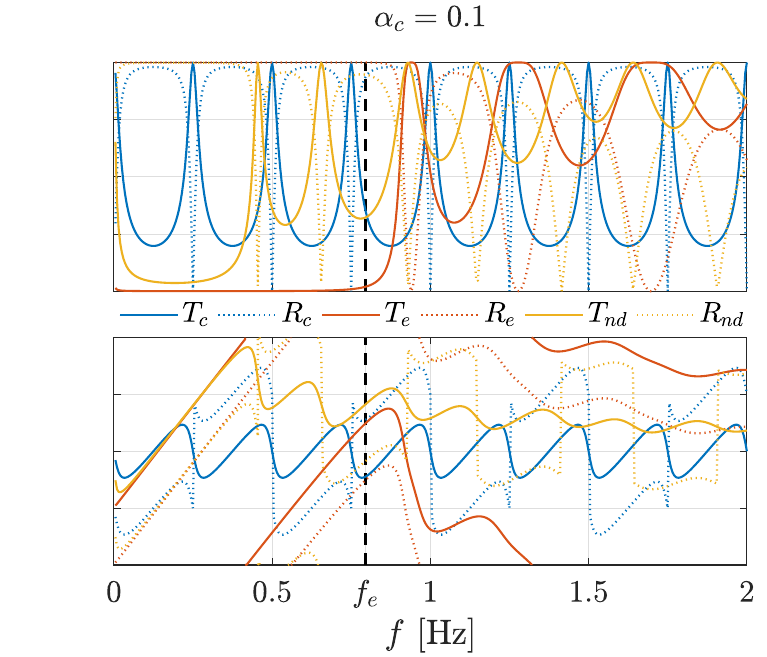}\label{fig:T and R slab-c}}
    \caption{The impedance profiles depicted in (a) are associated with the transmission and reflection coefficients shown in (b) for $\alpha_\text{c}=0.1$ and in (c) for $\alpha_\text{c}=0.2$.}
    \label{fig:alpha T R 1D}
\end{figure*}

\replaced{To consider this, we numerically assess three profiles that reduce the mass of a lens:}{
We compare three impedance profiles that all reduce the average properties of the lens by the same amount. This provides a measure of how much device feasibility improves. The considered profiles are:}
(i) constant $\tilde\alpha_{\text c}(x)\equiv\alpha_\text{c}\in\R$, (ii) exponential $\tilde\alpha_{\text e}(x)\coloneqq\alpha_{\text e}(|x|)$ and (iii) non-dispersive $\tilde\alpha_{\text{nd}}(x)\coloneqq\alpha_{\text{nd}}(|x|)$, where the argument $|x|$ represents the absolute value of $x$ used to have a symmetric lens.
The three profiles are \replaced{chosen}{selected} so that
\begin{equation}
    \begin{dcases}
        \tilde\alpha_{\text{nd}}(\pm R) = \tilde\alpha_{\text e}(\pm R) = 1
        \\
        \frac{1}{2R}\int_{-R}^R \tilde\alpha_{\text{nd}}(x) \,\text dx =
        \frac{1}{2R}\int_{-R}^R \tilde\alpha_{\text e}(x) \,\text dx =
        \alpha_{\text c}
    \end{dcases},
\end{equation}
the former condition to match the impedance between the lens and the background, the latter to have the same \replaced{mass of the lens}{average reduction of the lens properties} among the three cases.
\added{Although both the exponential and non-dispersive impedance profiles are constructed to start from 1--matching the impedance of the surrounding medium at the boundary--this does not imply that a fully impedance-matched device is considered attainable. In fact, due to the spatial variation of the refractive index within the lens, achieving impedance matching throughout the entire device is generally not possible. The impedance profiles presented in this work are therefore not intended as optimized designs, but rather as illustrative examples used to explore how different grading strategies affect transmission and dispersion.}
We choose the two configurations $\alpha_\text{c}=0.1$ and $\alpha_\text{c}=0.2$, and get the coefficients shown in Table~\ref{tab:coeff} and the \replaced{mismatches}{impedance profiles} shown in Figure~\ref{fig:alpha 1D}.
\added{
For $\alpha = 0.2$, the impedance at the center of the lens drops to about \qty{1}{\percent} of the ambient value for the exponential profile, and to \qty{4}{\percent} for the non-dispersive profile, which naturally leads to a smaller impedance jump. Such levels of mismatch are comparable to the challenge of coupling piezoelectric transducers to water \cite{pedersen1982impedance,li2017broadband}. In underwater acoustics, metamaterials have been developed with matched sound speeds and impedance values ranging from water-like \cite{brambilla2024high} to as low as \qty{1}{\percent} of it \cite{li2019three}. Although reaching such impedance levels is challenging, incorporating a graded impedance profile adds a valuable degree of freedom in designing the lens microstructure.
}

\noindent  Numerical simulations in frequency domain are performed in \Comsol{}, where harmonic waves of unitary amplitude are radiated from $-\infty$ \added{such that their phase is null when measured at the origin. 
Perfectly Matched Layers (PMLs) are applied at $x=\pm 2R$ to truncate the computational domain and a background pressure is imposed outside the lens.} The reflected and transmitted waves are measured at $x=\mp R$, respectively.
Figure~\ref{fig:alpha T R 1D} shows the result in terms of amplitude and phase over the frequency range $f\in[\qty{0}{\hertz},\qty{2}{\hertz}]$, where \replaced{$R=1$ and $c_0=1$}{$R=\qty{1}{\meter}$ and $c_0=\qty{1}{\meter\per\s}$} have been chosen.
\added{The phases are reported as if the signal were measured at the origin $x=0$ to simplify reading the plots: the desired behavior has transmission with unitary amplitude and null phase, and reflection with null amplitude. In this case, the signal would be transmitted as for the ideal impedance-matched lens.
The response near \qty{0}{\hertz} is reported for completeness, even though most lenses are designed in the high-frequency approximation. This is because our method applies to the full wave equation, and is relevant to devices that have been shown to operate in the full wave regime (e.g., the Maxwell's fish eye \cite{leonhardt2009perfect} or the optical black hole \cite{narimanov2009optical}). It also applies to lenses whose refractive index profile is derived using other approaches, such as coordinate transformation.
\\
Let us focus on Figure~\ref{fig:T and R slab-b}, where the average mismatch is $\alpha_\text{c}=0.2$.}
The constant profile $\tilde\alpha_\text{c}$ has a frequency-periodic behavior, with perfect transmission ($T\to1,\,R\to0$) at $f_\text{c}=\frac{n\,c_0}{4R},\,\forall n\in\mathbb N$, \added{as it is well known in the literature \cite{pedersen1982impedance}.}
\replaced{Conversely, we distinguish three regions for the exponential and the non-dispersive profiles.
At low frequency, both transmit a few energy because of the discontinuities at $x=\pm R$ and $x=0$.
In addition, the exponential profile transmit less below the minimum pass frequency $f_\text{e}\coloneqq\frac{\omega_\text{e}}{2\pi}$. This also affects transmission at mid frequency, where the behavior is similar for both profiles.
At higher frequencies, the amplitude $T$ and $R$ of both profiles tends to perfect transmission ($T\to1,\,R\to0$). Finally, an important difference is in the phase where the non-dispersive profile performs better.
These effects are accentuated when $\alpha_\text{c}=0.1$, because $f_{\text{e}}$ decreases further, as shown in Figure~\ref{fig:T and R slab-c}.}{
The exponential and the non-dispersive profiles have an oscillating behavior with perfect transmission at discrete frequencies, and at high frequencies in accordance with~\eqref{eq:generalized TR}. Moreover, except the perfect transmission at \qty{0}{\hertz}, their transmission is remarkably low till a first peak at \qty{0.64}{\hertz}
for the exponential profile and at \qty{0.43}{\hertz}
for the non-dispersive profile.
\\
For $\alpha_\text{c} = 0.2$, the cutoff frequency of the non-dispersive profile is slightly lower than that of the exponential profile, and the transmission is marginally higher for $\tilde\alpha_\text{nd}$ compared to $\tilde\alpha_\text{e}$--though the difference remains modest across the entire spectrum. This contrast becomes much more pronounced when $\alpha_\text{c} = 0.1$: in this case, the two peaks are at \qty{0.94}{\hertz} (exponential) and \qty{0.46}{\hertz} (non-dispersive), and the non-dispersive profile clearly exhibits a stronger transmission.
It is also worth noting that the non-dispersive peak remains nearly constant with respect to changes in the average mismatch $\alpha_\text{c}$, whereas $f_\text{e}$, representative of the first peak of $T_\text{e}$, varies significantly, consistent with the behavior predicted by Eq.~\eqref{eq:exp disp}.
\\
Except for the perfect transmission at zero frequency seen in Figure~2, a low-frequency gap is unavoidable when using a non-constant impedance profile, for two reasons. First, Eq.~(23) shows that total reflection occurs as $\omega \to 0$. Second, energy transmission is inefficient due to the dispersive nature of the exponential profile, the integral term in the non-dispersive profile, or both in a general scenario. These effects arise because the impedance variation appears abrupt to long-wavelength signals, no matter how gradual the transition. A constant profile performs better in this regime; however, the position of the first transmission peak depends only on the lens size, not on the value of $ \alpha_\text{c}$.

Finally, we examine the phase plots to assess dispersion.
The constant-mismatch profile exhibits a phase that is periodic in frequency, reflecting the resonances occurring within the lens \cite{kossoff1966effects}. In contrast, the graded profiles show a peak in the phase plot in correspondence of their respective amplitude peaks, and their phases approach zero at higher frequencies. Notably, the non-dispersive profile exhibits significantly less phase distortion than the exponential profile, indicating lower dispersion in the transmitted signal—consistent with the design goals. The non-dispersive profile introduces dispersion only at the interfaces, whereas the exponential profile is dispersive also throughout the volume. Since $\tilde\alpha_\text{nd}$ has more pronounced slope discontinuities at the interfaces, the severe dispersion observed with $\tilde\alpha_\text{e}$ must therefore be attributed to volume dispersion around the frequency $f_{\text{e}}$.
\\
Both the results in terms of amplitude and phase make the non-dispersive profile a valuable candidate when a strong mismatch inside the lens is unavoidable.
Moreover, the constants $a_i$--which we selected to satisfy $\alpha(R) = 1$ and $ \alpha_\text{c} = 0.1, 0.2$--could be chosen differently to meet specific application requirements, such as enhancing transmission in selected frequency bands. This flexibility may provide some control over the fluctuations observed in Figure~\ref{fig:alpha T R 1D}, although they are common to any impedance profile due to resonances arising from mismatch within the lens. A systematic optimization of these parameters is beyond the scope of this work.
}

\begin{figure}
    \centering
    \includegraphics[width=240pt]{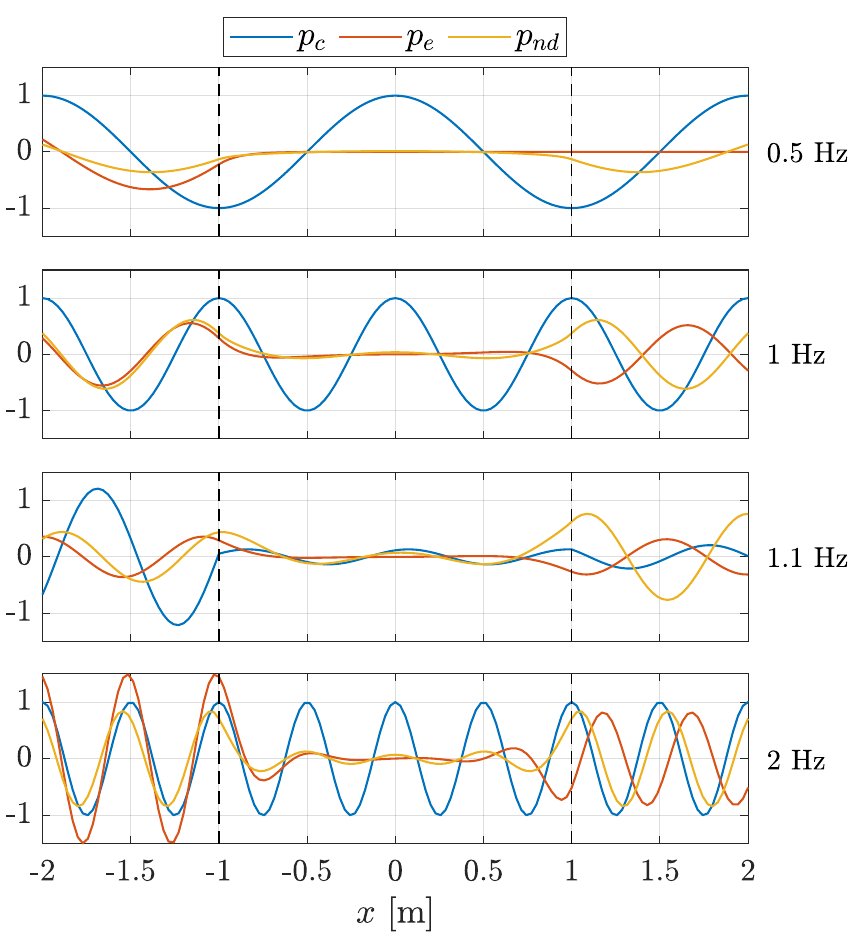}
    \caption{Total pressure field resulting from a harmonic wave of frequency \qtylist{0.5;1;1.1;2}{\hertz}, radiated from $-\infty$ and interacting with the lens located in the interval $[ \qty{-1}{\meter}, \qty{1}{\meter} ]$.}
    \label{fig:pressure 1D}
\end{figure}

\added{
Figure~\ref{fig:pressure 1D} provides further insight by showing the pressure field at frequencies \qtylist{0.5;1;1.1;2}{\hertz} for an average impedance mismatch of $\alpha_\text{c} = 0.1$.
Perfect transmission is observed for the constant impedance profile at \qtylist{0.5;1;2}{\hertz}, whereas a sharp drop in performance occurs at \qty{1.1}{\hertz}, indicating narrowband behavior around \qty{1}{\hertz}.
At \qty{0.5}{\hertz}, the exponential profile fails to transmit any signal, while the non-dispersive profile allows only weak transmission.
At \qtylist{1;2}{\hertz}, the wave transmitted by the non-dispersive profile closely matches that of the constant profile, approaching perfect transmission.
Note that for the non-constant profiles, the pressure amplitude inside the lens remains significantly lower than $1$, even when transmission is efficient. This is due to the local rescaling caused by the spatial variation of properties.
}

\section{L\"uneburg lens with distributed mismatch}
\replaced{
Let us now consider a 2D lens, where dispersion may play an important role on the interference between waves traveling along different paths.
In the following example, we also introduce a refractive index that varies along the radius, as is typical for circular lenses.
}{
We now extend our analysis to a two-dimensional (2D) configuration, where wave propagation becomes more complex and dispersion significantly affects interference of rays traveling along different paths. Studying dispersion in one dimension is relatively straightforward as it reduces to computing the phase shift along a single path—that is, determining the dispersion relation as we did for the exponential profile. In contrast, dispersion analysis in 2D is more involved, especially due to the refractive index profile of a lens. However, our approach allows us to derive a non-dispersive impedance profile without explicitly computing the 2D dispersion relation.
The following example considers a circular lens with a radially varying refractive index to examine how impedance mismatch and dispersion affect wave focusing in a more realistic and practically relevant geometry.
}\\
An impedance profile that limits dispersion is found by a similar procedure as before, for an axisymmetric lens with radius $R$ centered at the origin. Let $n(r)$ be the refractive index profile of the lens, where $r$ is the distance from its center. Such a lens is perfectly matched if
\begin{align}\label{eq:match lens prop}
    \rho &= \rho_0\,n(r),
    &
    K &= K_0/n(r),
\end{align}
so that $Z=Z_0$ everywhere. Note that $\rho$ and $K$ are bounded and non-null if $n$ is bounded and non-null.
Let $\alpha(r)$ be an arbitrary impedance profile along the radius, the wave equation for the mismatched lens reads as:
\begin{equation}\label{eq:wave 2D}
    \nabla\cdot\Big(\frac{1}{\alpha\, n\rho_0}\nabla p\Big)= \frac{n}{\alpha\, K_0} p_{tt},
\end{equation}
Let us assume that $p_1$ is the solution to~\eqref{eq:wave 2D} when $\alpha(r)\equiv 1$.
We assume that the general \deleted{the} solution $p_\alpha$ has the form
\begin{equation}
    p_\alpha = G(r,p_1),
\end{equation}
where the function $G(r,p)$ explicitly depends on $r$ and not on the tangential coordinate to enforce axial symmetry. Substituting $p_\alpha$ into \eqref{eq:wave 2D} and repeating the same calculations as before but in polar coordinates, we obtain
\begin{subequations}\label{eq:def fgh}
\begin{align}
    G(r,p) &= g(r)\,p + h(r)
    \\
    \alpha_\text{nd}(r) &= a_{0,\text{nd}} \big( a_{1,\text{nd}} + N(r) \big)^{-2}
    \\
    g(r) &= a_{2,\text{nd}} \big( a_{1,\text{nd}} + N(r) \big)^{-1}
    \\
    h(r) &= a_{3,\text{nd}} \big( a_{1,\text{nd}} + N(r) \big)^{-1}+a_{4,\text{nd}}
\end{align}
\end{subequations}
for some constants $a_{i,\text{nd}}$, $i\in\{0,1,2,3,4\}$, and where $N(r) \coloneqq \int_0^r n(\tilde r)/\tilde r\,\text d\tilde r$.
Then, for any choice of the five integration constants, the pressure field $p_1$ is modified without introducing material dispersion.
The constant mismatch is recovered in the limit of $a_{0,\text{nd}}\propto a_{1,\text{nd}}^2\to\infty$. \\
Note that for $r\to0^+$, $N$ is unbounded if $n\neq0$.
This happens when the phase velocity is finite, as is typical for GRIN lenses.
So $\alpha(r\to0^+)=0$ regardless of the values of $a_0$ and $a_1$. This implies that vanishing properties are required in the center of the lens.
This condition poses no issue when a sensor or an object must be placed at the center, as in sensing~ \cite{narimanov2009optical,cominelli2024optimal} or cloaking~ \cite{chen2017broadband,quadrelli2021experimental} devices. In cases where this is not feasible, introducing a small void region at the center is sufficient to satisfy this requirement, at least in acoustic applications.

As an example, we consider the circular L\"uneburg lens that focuses a set of parallel rays into one point on the external circumference~ \cite{luneburg1966mathematical}.
The refraction index profile is
\begin{align}\label{eq:n luneb}
    n_\text{L}(r) &= \sqrt{2 - \frac{r^2}{R^2}}, & r\in [0, R]
\end{align}
The properties of the impedance-matched lens are defined according to \eqref{eq:match lens prop}.
\replaced{We aim to reduce the mass}{As before, we aim to rescale the average properties} of the lens \replaced{$m\coloneqq\int_{\text{lens}}\rho\,\text dV$}{$\langle\alpha\rangle\coloneqq\int_{\text{lens}}\alpha\,\text dV/\int_{\text{lens}}\text dV$} while maintaining the imaging characteristics. The \deleted{mass of the matched lens $m_{\text{ref}}$ and its} performance \replaced{are}{of the matched lens is} taken as reference and three strategies are compared: (i) the constant impedance profile $\tilde\alpha_\text{c}\equiv\alpha_{\text c}\in\R$, (ii) the exponential profile $\tilde\alpha_{\text e}=a_{0,\text e}\,e^{a_{1,\text e} r}$, and (iii) the non-dispersive profile $\tilde\alpha_{\text{nd}}=a_{0,\text{nd}}(a_{1,\text{nd}}+N_\text{L}(r))^{-2}$, where
\begin{equation}
    N_\text{L}(r) =
    n_\text{L}(r) + \sqrt2\,\atanh\Big(n_\text{L}(r)/\sqrt2\Big),
\end{equation}
and $\atanh$ denotes the inverse of the hyperbolic tangent.
The constants $\alpha_{\text c}$, $a_{0,\text e}$, $a_{1,\text e}$, $a_{0,\text{nd}}$, $a_{1,\text{nd}}$ are defined such that
\begin{subequations}\label{eq:vincoli}
\begin{gather}
    \langle\tilde\alpha_{\text c}\rangle = \langle\tilde\alpha_{\text e}\rangle = \langle\tilde\alpha_{\text{nd}}\rangle = \alpha_\text{c},
    \\
    \tilde\alpha_{\text e}(R) = \tilde\alpha_{\text{nd}}(R) = 1.
\end{gather}
\end{subequations}
The former condition is such that the \replaced{mass}{average properties} of the lens \replaced{is}{are} reduced by the same amount $\alpha_\text{c}\in\{0.1,0.2\}$; the latter requires the lens to match the background on the interface for $\alpha_{\text e}$ and $\alpha_{\text{nd}}$. Eq.~\eqref{eq:vincoli} is solved numerically to find the coefficients shown in Table~\ref{tab:coeff 2D}. The minimum frequency $f_\text{e}$ of the exponential profile is calculated using the formula obtained in 1D to have a reference.

\begin{table}[]
    \centering
    $\begin{array}{c|ccc|cc}
    \alpha_{\text{c}} \,[-] & a_{0,\text e}\,[-]  & a_{1,\text e}R\,[-] & f_{\text{e}} \,[Hz] &a_{0,\text{nd}} \,[-]& a_{1,\text{nd}} \,[-]
    \\    \midrule
    0.1 & \num{9.3e-8} & 16.2 & 1.29      & \num{6.7e-3} & 0.16
    \\
    0.2 & \num{4.1e-4} & 7.80 & 0.62      & \num{4.1e-2} & 0.043
    \end{array}$
    \caption{Coefficients of the 2D impedance profiles.}
    \label{tab:coeff 2D}
\end{table}
\begin{figure*}
    \centering
    \subfloat[]{\includegraphics[scale=0.45]{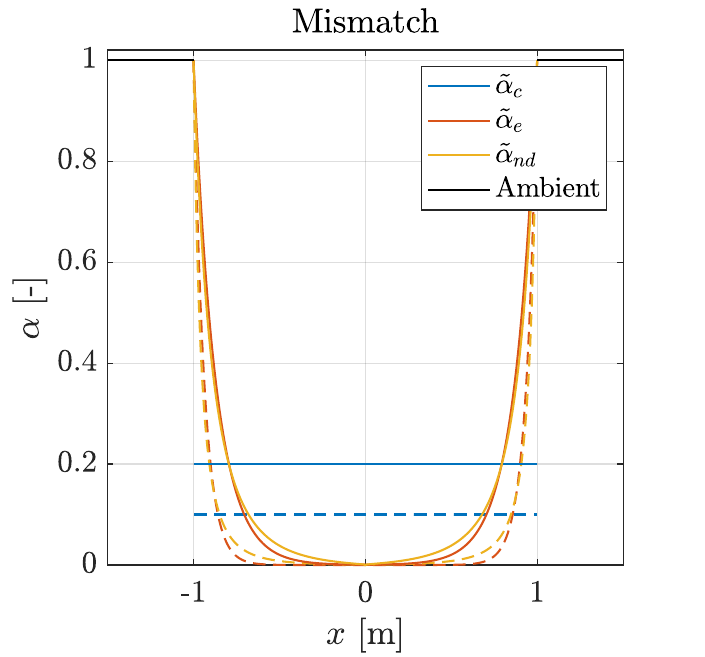}\label{fig:alpha 2D}}
    \subfloat[]{\includegraphics[scale=0.46]{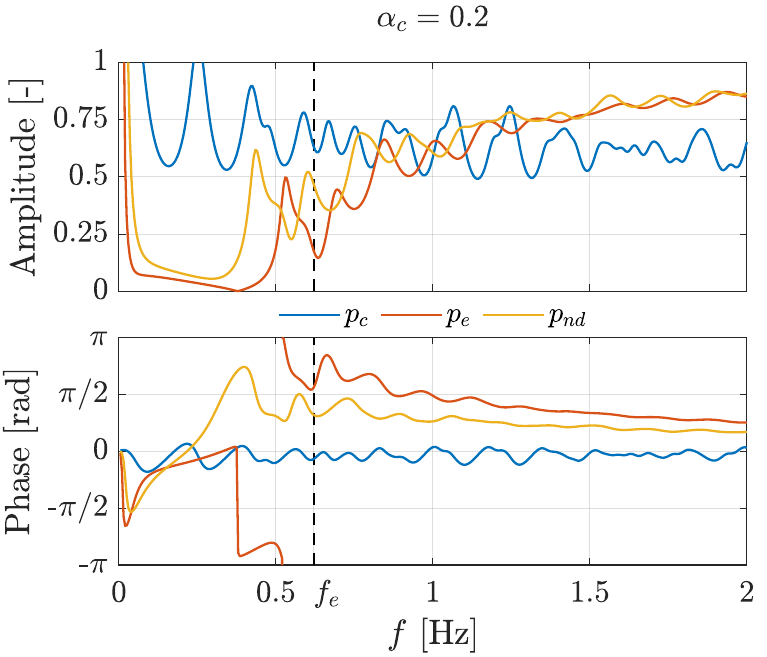}\label{fig:Fig4b}}
    \subfloat[]{\includegraphics[trim=30 0 0 0,clip,scale=0.46]{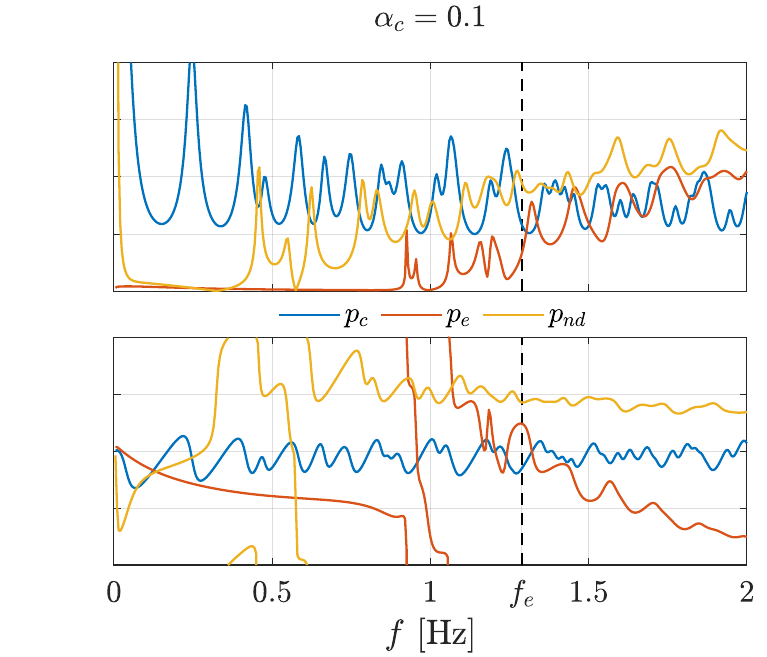}\label{fig:Fig4c}}
    \caption{The impedance profiles shown in (a) are associated with the pressure measured in (b) for $\alpha_\text{c} = 0.1$ and in (c) for $\alpha_\text{c} = 0.2$.}
    \label{fig:p_focus Luneburg}
\end{figure*}

The performance of the impedance profiles is evaluated numerically. A plane wave of unitary amplitude propagating in the $x$ direction is emitted from $(-\infty,0)$ and focused by the lens at the point $F=(R,0)$ on the outer surface of the lens.
\added{The simulation has been performed in \Comsol{} imposing a background pressure field outside the lense and using PMLs to truncate the computational domain.}
Figure~\ref{fig:p_focus Luneburg} shows the \added{total} pressure measured in $F$ for the three configurations, normalized with respect to the measurement by the matched lens
\added{so that the desired behavior is $|p|=1$ and $\angle p = 0$.}
\\
\replaced{By comparing Figures~\ref{fig:Fig4b} and \ref{fig:Fig4c}, we note that the constant mismatch affects the wave field similarly across the entire spectrum, while three distinct frequency regions can be identified for the exponential and non-dispersive profiles
At low frequencies, the smooth impedance profiles have a greater effect on performance than the constant profile. This is because at long wavelengths, the gradual transitions in material properties are perceived as a jump deeper than for a constant mismatch. At high frequencies, both exponential and non-dispersive impedance profiles do not noticeably change the performance of the lens because there are negligible reflections at the lens-background interface and dispersion within the lens. More importantly, in the mid-frequency range--where the wavelength is comparable to the size of the lens--the dispersion introduced by the exponential profile becomes significant, while the non-dispersive profile shows better performance.
}{Let us consider Figures~\ref{fig:Fig4b} and \ref{fig:Fig4c}. When the constant mismatch profile is adopted, the response is characterized by resonant peaks similar to those of the 1D configuration. The amplitude of these peaks decreases as the frequency increases, until the pressure amplitude $|p_\text{c}|$ eventually oscillates around a constant value. This value is higher for a higher $\alpha_\text{c}$.\\
In contrast, the exponential and non-dispersive profiles exhibit a cutoff frequency below which little to no energy is focused at point $F$. For the exponential profile, this cutoff is roughly estimated by the 1D formula for $f_\text{e}$, and it varies significantly with $\alpha_\text{c}$; for the non-dispersive profile, the cutoff is only slightly affected by $\alpha_\text{c}$ and lower than $f_\text{e}$.
Above their cutoff frequency, both profiles exhibit amplitude oscillations that gradually diminish as the frequency increases. Moreover, the average values of $|p_\text{e}|$ and $|p_\text{nd}|$ monotonically increase toward the theoretical perfect transmission at high frequencies. On average, $|p_\text{nd}|$ is higher than $|p_\text{e}|$ (especially for $\alpha_\text{c}=0.1$), and from approximately \qty{1}{\hertz} onward, it also exceeds $|p_\text{c}|$.
Regarding phase distortion--used here as a measure of dispersion--it is significantly lower for the non-dispersive profile compared to the exponential one, but it remains higher than that of the constant mismatch profile. Nonetheless, it tends to zero as frequency increases.}

\begin{figure*}  
    \centering
    \subfloat[]{\includegraphics[width=0.3\linewidth,trim=0 10 0 0]{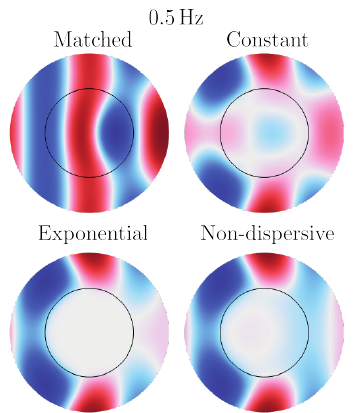}} 
    \subfloat[]{\includegraphics[width=0.3\linewidth,trim=0 10 0 0]{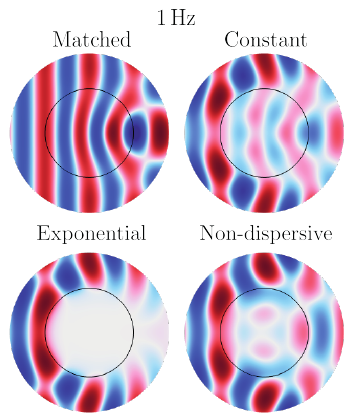}} 
    \subfloat[]{\includegraphics[width=0.3\linewidth,trim=0 10 0 0]{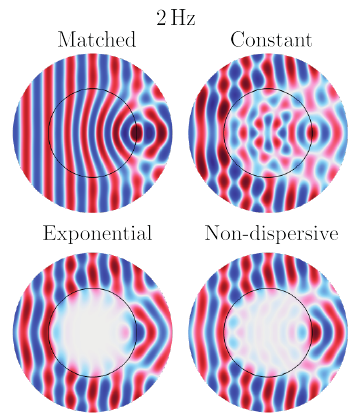}} 
    \includegraphics[height=0.315\linewidth,trim=0 13 0 0]{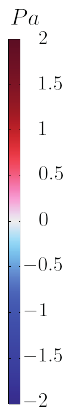} 
    \\
    \subfloat[]{\includegraphics[width=0.3\linewidth,trim=0 10 0 0]{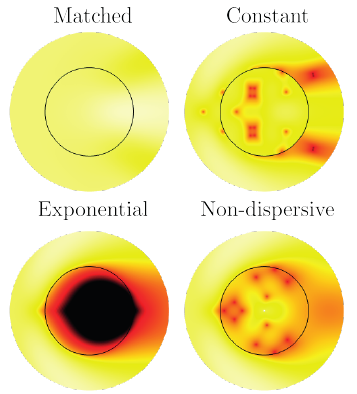}} 
    \subfloat[]{\includegraphics[width=0.3\linewidth,trim=0 10 0 0]{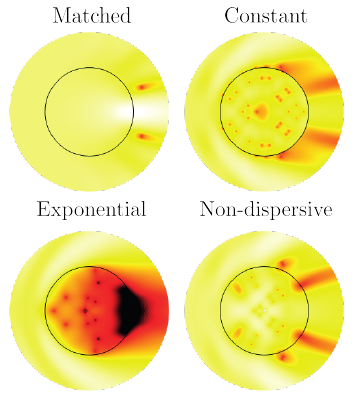}} 
    \subfloat[]{\includegraphics[width=0.3\linewidth,trim=0 10 0 0]{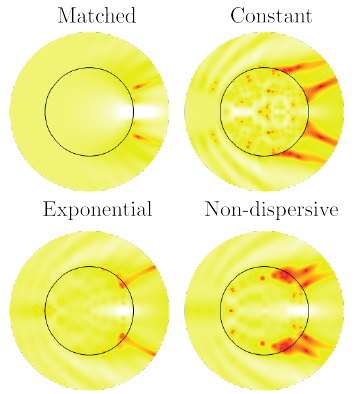}} 
    \includegraphics[height=0.315\linewidth,trim=0 13 0 0]{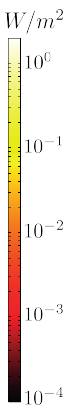} 
    \caption{Pressure (a–c) and intensity (d–f) fields of the L\"uneburg lens with $ \alpha_\text{c} = 0.1$. The black contour outlines the lens, and the computational domain is enclosed by PMLs, which are not shown. For improved readability, the intensity color scale is clipped below a minimum value of \num{1e-4}.}
    \label{fig:2d field}
\end{figure*}

\added{
Figure~\ref{fig:2d field} shows the pressure and intensity fields for the matched lens and the three mismatched profiles (constant, exponential, and non-dispersive) at \qtylist{0.5;1;2}{\hertz}, with $\alpha_\text{c} = 0.1$.
In the matched case, energy is efficiently focused at the expected focal point to the right of the lens, especially at \qty{1}{\hertz} and \qty{2}{\hertz}, confirming the lens ideal behavior. The constant impedance profile performs similarly across all three frequencies, demonstrating relatively stable focusing performance.
\\
In contrast, the exponential profile fails to concentrate energy effectively at \qtylist{0.5;1}{\hertz}, likely due to increased reflection and dispersion. Both the exponential and non-dispersive profiles show very low pressure amplitudes inside the lens, which may suggest inefficient transmission. However, the intensity fields of the non-dispersive profile reveal that energy is still efficiently carried through the lens, particularly at \qty{2}{\hertz}. This apparent contradiction is due to the significant reduction in material property magnitudes near the lens center, which affects the pressure but not the intensity.
Overall, the non-dispersive profile maintains good energy transport, with intensity levels at the focal region comparable to those of the matched lens, especially at higher frequencies.
}

\added{In principle, the framework we presented can be extended to anisotropic media. If the target lens is anisotropic--e.g., with a tensorial mass density--our approach can still be applied by rescaling both the density tensor and the stiffness by a scalar function $\alpha$. Then, one can substitute the assumed pressure field (proportional to a solution of the matched anisotropic lens) into the governing acoustic equations. This yields a modified set of coupled differential equations for $\alpha$, similar in spirit to \eqref{eq:diff 1D 456}, although the resulting system may be more complex.\\
Many anisotropic acoustic devices are designed using transformation acoustics, starting from a virtual isotropic space. In this context, our impedance profiling strategy can be applied directly in the virtual (isotropic) domain. The resulting non-dispersive impedance variation is then naturally mapped into anisotropic physical parameters through the coordinate transformation. We have recently followed this approach in the context of underwater acoustic cloaking \cite{quadrelli2025reduced}.
While a full treatment of anisotropy is beyond the scope of this work, we believe that incorporating controlled impedance mismatches into more general transformation-based or anisotropic designs is promising.}

\section{Conclusion}

In this paper, we present a method for grading the impedance of lenses, providing a degree of freedom in their design. This approach is particularly valuable in scenarios where it is difficult to achieve material properties similar to those of the surrounding medium, as is often the case in acoustics.
We discuss phase distortion and highlight that it generally occurs either at the device interfaces or within its volume.
We quantify these two phenomena in the commonly used exponential profile and derive a generalized formula for the transmission and reflection problem \added{for 1D normal incidence}. Our analysis shows that dispersion arises from abrupt changes in the slope of material properties.
We construct a family of impedance profiles that prevent dispersion within a device, thereby improving its performance \added{and/or relaxing the design}.
\\
This approach is extended to 2D axisymmetric lenses where the impedance varies continuously along the radial direction. As an illustrative example, we apply the method to the design of a L\"uneburg lens and demonstrate the practical potential of the approach.
While the present work focuses on acoustics, the proposed method is broadly applicable to other physical systems governed by similar mathematical structures.

\begin{acknowledgments}
I thank A.\ Cominelli for the insightful discussions we had that enriched this work.
\end{acknowledgments}

\section*{Conflict of interest}
The author declares that he has no conflicts of interest.

\section*{Data availability}
The codes used in this work are available at  \url{https://github.com/SebaComi/Graded_impedance}.

\bibliography{bibliography}

\end{document}